%% file: iscience_reverbreview.tex
\documentclass[review]{elsarticle}


\biboptions{authoryear}

\usepackage[table]{xcolor}
\usepackage{graphicx, sidecap}
\usepackage{natbib}
\usepackage{natbibspacing}
\usepackage{rotating}
\usepackage{pdfpages}
\usepackage{hyperref}
\usepackage{amssymb}
\usepackage{relsize}

\newcommand\xmm{{\it XMM-Newton}}
\newcommand{\nicer}{\textit{NICER}}
\newcommand{\nustar}{\textit{NuSTAR}}
\newcommand{\hst}{\textit{HST}}
\newcommand{\rl}{$R_{\rm BLR}-L_{\rm AGN}$}
\newcommand{\msigma}{$M_{\rm BH}-\sigma_{\star}$}

\newcommand{\ion}[2]{#1$\;${\small\rmfamily{#2}}\relax}

\begin{document}

\begin{frontmatter}

\title{Reverberation mapping of Active Galactic Nuclei: from X-ray corona to dusty torus}

\author{Edward M. Cackett\corref{mycorrespondingauthor}}
\cortext[mycorrespondingauthor]{Corresponding author}
\ead{ecackett@wayne.edu}
\address{Department of Physics \& Astronomy, Wayne State University, 666 W. Hancock St, Detroit, MI 48201, USA}

\author{Misty C. Bentz}
\address{Department of Physics \& Astronomy, Georgia State University, Atlanta, GA 30303, USA}

\author{Erin Kara}
\address{MIT Kavli Institute for Astrophysics and Space Research, Cambridge, MA 02139, USA}

\begin{abstract}

The central engines of Active Galactic Nuclei (AGNs) are powered by accreting supermassive black holes, and while AGNs are known to play an important role in galaxy evolution, the key physical processes occur on scales that are too small to be resolved spatially (aside from a few exceptional cases). Reverberation mapping is a powerful technique that overcomes this limitation by using echoes of light to determine the geometry and kinematics of the central regions. Variable ionizing radiation from close to the black hole drives correlated variability in surrounding gas/dust, but with a time delay due to the light travel time between the regions,  allowing reverberation mapping to effectively replace spatial resolution with time resolution. Reverberation mapping is used to measure black hole masses and to probe the innermost X-ray emitting region, the UV/optical accretion disk, the broad emission line region and the dusty torus. In this article we provide an overview of the technique and its varied applications. 

\end{abstract}

\end{frontmatter}

\section{Introduction}

\begin{figure}
\centering
\includegraphics[width=\textwidth]{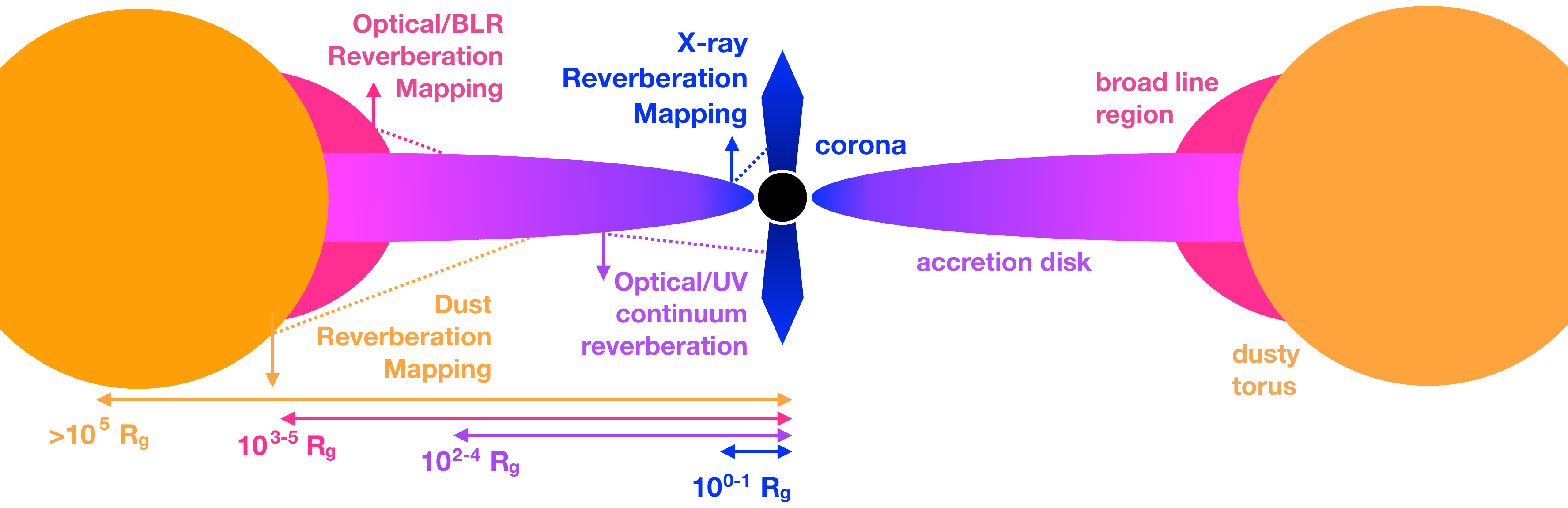}
\caption{A cross-sectional schematic of an AGN, highlighting the main components and the four types of reverberation discussed in this review, X-ray reverberation (Section~\ref{sec:xray}), optical/UV continuum reverberation (Section~\ref{sec:disk}), broad line region reverberation (Section~\ref{sec:blr}), and dust reverberation (Section~\ref{sec:dust}) with general radial scales from the black hole indicated by labels.}
\label{fig:reverb_schematic}
\end{figure}

It is now widely accepted that most massive galaxies contain a supermassive black hole (SMBH) at their center. The masses of these SMBHs correlate with properties of the galaxy itself, such as the mass of the bulge and stellar velocity dispersion \citep{magorrian98,gebhardt00}, evincing a link between the growth of the galaxy and the SMBH. A small fraction of galaxies are known as active galaxies and host Active Galactic Nuclei (AGNs), in which strong electromagnetic radiation is emitted across the entire spectrum from their central region, emission that is powered by the release of gravitational potential energy as material accretes onto the SMBH.  This accretion can drive winds and outflows that influence the AGN host galaxy and explain the link between galaxy and SMBH properties \citep{silk98,fabian99}. This important process in galaxy evolution is often referred to as `AGN feedback' \citep[see][for a review]{fabian12}. Understanding the process of accretion and the geometry and structure of gas in the inner regions of AGNs is therefore an important piece in understanding galaxy evolution. A simplified sketch of the inner regions and basic structure of an AGN is shown in Fig.~\ref{fig:reverb_schematic}, showing the accretion disk, X-ray corona, broad emission line regions, or BLR, and dusty torus \citep[see][and references therein for a detailed review of AGNs]{netzer15}.

The angular size of the vast majority of AGNs is too small to be spatially resolved with current techniques. For example, if we consider a relatively nearby AGN at a distance of 50 Mpc, with a black hole mass of $M = 10^7$ M$_\odot$, a size scale of 10,000 $R_{\rm G}$ (similar to the size scale of the BLR; $R_{\rm G} = GM/c^2$) corresponds to an angular size of approximately $2\times10^{-5}$ arcsec. There are a few limited, and notable, cases where the AGN is near enough and the black hole massive enough that the angular size is within reach of some methods.  For instance, the Event Horizon Telescope used interferometry at sub-mm wavelengths to image the black hole shadow in M87 \citep{EHT_M87} and the GRAVITY interferometer on the VLT has also been able to resolve the BLR in 3C~273 and IRAS~09149-6206 \citep{gravity_3c273,gravity_iras09149}.  Whereas size scales of $\sim10^5~R_{\rm G}$ are currently inaccessible to all but a few AGN, in the future, X-ray interferometry may eventually make it possible to resolve scales of a few $R_{\rm G}$ in the nearest Seyfert galaxies \citep{uttley19}. However, the vast majority of AGNs remain out of reach of current interferometric techniques, thus, other methods are needed to study the innermost regions of AGNs.  

The focus of this review is reverberation mapping, a technique that swaps spatial resolution for time resolution, allowing black hole mass measurements and the determination of the size scales of emitting regions associated with X-rays, the UV/optical continuum, BLR, and dusty torus.  Measuring the size and structure of the BLR was the first proposed application for reverberation mapping. \citet{bahcall72} recognized that time variability in the intensity of emission from a central source would affect the observed intensity of emission from photoionized gas in active galaxies and some novae.  Building on this, \citet{blandford82} developed a framework for inverting the observed time-dependence of broad emission-line variations to map out the structure of the BLR.  In essence, time resolution can provide constraints on the emissivity and position of photoionized gas in a spatially-unresolved source, even when the source is located at a distance of millions to billions of parsecs.

\subsection{Reverberation Mapping Basics}

The basic principle of reverberation mapping is to measure the time delay, or lag, between flux variations in an ionizing source, and the flux variations from the surrounding region that is being irradiated.  Changes in the irradiating flux will drive changes in the reprocessed emission so light curves from the two regions should be strongly correlated.  However, since the light from the reprocessed emission has to travel an extra distance to get to us (see Fig.~\ref{fig:reverb_schematic}), the variations in the reprocessed emission will arrive later than the variations in the ionizing flux.  The lag, $\tau$, will depend on the exact geometry of the system, but the average lag will be of the order $R/c$, where $R$ is the typical radius of the emitting region.

The basic simplifying assumptions of reverberation mapping are that (a) the irradiating flux originates from a single central source, (b) that the light-travel time is the most important timescale, and that (c) the relationship between the observed reprocessed and ionizing fluxes is linear.  If we describe the ionizing and reprocessed light curves, $F_r(t)$ and $F_i(t)$ respectively, as consisting of a constant plus variable component, i.e. $F_i(t) = \bar{F}_i + \Delta F_i(t)$ and $F_r(t) = \bar{F}_r + \Delta F_r(t)$, then the relationship between the variable components of each light curve ($\Delta F_i(t)$ and $\Delta F_r(t)$) can be described by:
\begin{equation}
\Delta F_r(t) = \int_0^{\tau_{\rm max}} \Psi(\tau) \Delta F_i(t - \tau) d\tau 
\end{equation}
where  $\Psi(\tau)$ is the transfer function (or impulse response function) which encodes all the information about the geometry of the reprocessing region. In other words, the reprocessed light curve is a blurred, delayed version of the ionizing light curve, where $\Psi(\tau)$ can be thought of as the blurring kernel. This framework can be extended further to include velocity (or energy) dependence to the light curves and transfer function, allowing for a study of the kinematics of the reprocessing region also.

Reverberation mapping therefore requires measuring the flux variations of the AGN at different wavelengths.  From this, we first aim to recover the lag between the light curves since this gives the responsivity-weighted radius of the reprocessing region. But, ultimately we would like to recover the transfer function in order to fully map out the geometry and dynamics of the region.  Methods employed to measure the lag depend on the nature of the observed light curve.  For instance, optical light curves obtained from the ground are never continuous, with poor weather causing gaps.  This necessitates a time domain approach to measuring the lag.  In X-rays, however, the relevant timescales are short enough that continuous light curves can often be obtained, allowing for Fourier analysis techniques. We briefly describe both approaches below.

Commonly, measuring lags in the time domain uses the cross-correlation function. Essentially one shifts the light curves in time with respect to each other and measures the degree of correlation at each lag. With non-continuous light curves, the question becomes how to deal with the gaps.  Often linear interpolation is used to fill between the gaps \citep{gaskell86,white94,peterson04}, or alternatively interpolation can be avoided by using the discrete correlation function  \citep{edelson88,white94}.  One advantage of the cross-correlation approaches are that they are simple, making no assumptions, and rely on the data alone.  However, when the data are not well-sampled they do not perform well.  More recently, Markov Chain Monte Carlo (MCMC) methods that assume the underlying AGN variability can be described by a damped random walk or by Gaussian processes, have been developed to fit the light curves and determine the lag \citep{Zu11,Zu16,starkey16, Yu20}.  The advantage of these methods is that with some assumptions about the variability characteristics and transfer function shape they can provide lag estimates with less well-sampled data than traditional cross correlation methods, with the disadvantage being the dependence on the validity of the assumptions made. 

The Fourier analysis techniques employed in X-ray reverberation analysis is described in detail in \citet{uttley14}. Briefly, the cross-spectrum of the light curves in two bands is used to determine the lag at each Fourier frequency, and is calculated as follows.  If the light curve in one energy range is $s(t)$ and its corresponding Fourier transform is $S(f)$, and in the other band the light curve and Fourier transform are $h(t)$ and $H(f)$, the cross-spectrum is calculated via $C(f) = S^*(f) H(f)$ where $S^*(f)$ is the complex conjugate of $S(f)$. The phase of the cross-spectrum, $\phi(f)$, gives the phase lag at each Fourier frequency.  This can be converted to a time lag, $\tau(f) = \phi(f)/2\pi f$. In practice the lags are averaged over a range of Fourier-frequencies, and/or a number of light curve segments.  This technique therefore gives the lag as a function of Fourier frequency, and so can separate lags from different processes occurring on different timescales. Not all X-ray telescopes provide long, continuous light curves where the approach can be directly applied, and so there is a need to also be able to determine Fourier-resolved lags from data containing gaps.  Several methods have been developed using maximum likelihood fitting and MCMC approaches \citep{zoghbi13b, wilkins19}.

Reverberation mapping was first used to measure the sizes of the BLR by measuring lags between the UV or optical continuum and broad emission lines such as \ion{C}{IV} and H$\beta$.  But, the principle has been extended to study the dusty torus, the accretion disk (UV/optical continuum) and the X-ray corona.  In this review we will outline the application of reverberation mapping to each of these different size-scales and structures around the SMBH (Fig.~\ref{fig:reverb_schematic}). The scale of interest is measured in terms of the gravitation radius, $R_{\rm G}$, and the equivalent light-crossing time is given by $t_c = R_{\rm G}/c$ (see Table~\ref{tab:sizes} for the light-crossing times). First, in Section \ref{sec:xray} we discuss the innermost X-ray-emitting part of the accretion flow at just a few to tens of $R_{\rm G}$.  Next (Section \ref{sec:disk}), we'll move outwards to discuss the UV and optical emitting continuum and accretion disk  at scales of 100 to 10000 $R_{\rm G}$. Moving further out still, in Section~\ref{sec:blr} we will discuss the broad emission line region at scales of thousands to tens of thousands of $R_{\rm G}$. Finally, Section~\ref{sec:dust} discusses the inner edge of the dust-emitting region at scales of tens of thousands to hundreds of thousands of $R_{\rm G}$ and beyond.  

\begin{table}
    \centering
    \begin{tabular}{rccc}
    \hline
    Region & Size scale & \multicolumn{2}{c}{Light-crossing time for a black hole mass of}  \\
           &  ($R_{\rm G}$) & $10^7 M_\odot$ & $10^9 M_\odot$ \\ \hline
    X-ray  &  10  & 500s & 14 hours \\
    UV/optical disk & $10^2 - 10^4$ & 0.06 -- 6 days & 6 -- 600 days \\
    BLR & $10^3 - 10^5$ & 0.6 -- 60 days & 60 -- 6000 days \\
    Dusty torus & $>10^5$  & $>$60 days & $>$6000 days \\ \hline
    \end{tabular}
    \caption{Order of magnitude size scales and their equivalent light-crossing times for the different regions of an AGN discussed in this review.}
    \label{tab:sizes}
\end{table}

\section{X-ray reverberation}  \label{sec:xray}

As gas funnels in towards the SMBH at the center of a galaxy, collisions and angular momentum conservation cause the formation of an optically thick, geometrically thin accretion disk around the black hole, classically described by \citet{SS73}. This disk is mildly ionized and produces weak magnetic fields that are responsible for the outward transfer of angular momentum, allowing for material to fall towards the black hole. In luminous, actively accreting AGN, material flows inwards through the accretion disk up to the innermost stable circular orbit (ISCO), beyond which the gas plunges towards the black hole on a ballistic trajectory\footnote{The situation is quite different in lower accretion rate AGN, where the density of the flow is so low that very little accretion energy is radiated away, thus heating up the inner flow, and effectively truncating the inner edge of the accretion disk at radii larger than the ISCO \citep{narayan96}.}.  The location of the ISCO is dictated by the spin of the black hole, as General Relativistic frame dragging effects will support more orbits closer to the black hole as the spin increases. X-ray reverberation aims to exploit this fact, and measure the inner edge of the accretion disk in luminous, radiatively efficient AGN, in order to measure a fundamental black hole parameter, its spin. 

The spin of the black hole also has important astrophysical implications, as the overall radiative efficiency of the accretion process is largely determined by the black hole spin (6\% for non-rotating black holes vs. 42\% for maximally spinning black holes). Moreover, the spin is a by-product of the underlying growth mechanism of the SMBH, as gas that accretes on to the black hole via a prograde accretion disk will eventually spin up the black hole \citep{Volonteri2003}. Randomly-oriented black hole-black hole mergers, on the other hand, will tend to spin-down the black hole. Therefore, if we can measure a distribution of black hole spins across the universe, we can understand the relative importance of the accretion process versus mergers in the growth of SMBHs \citep{bertivolonteri08}. 

The best probe of the ISCO in AGN is in the X-ray band. X-rays are ubiquitous in AGN systems and originate within a few gravitational radii of the black hole. The accretion disk is hot enough to emit thermal radiation that peaks in the Extreme Ultraviolet band (see schematic, Fig.~1). Some of those UV photons scatter off of mildly relativistic electrons in an region close to the black hole that is known as the corona \citep{haardt91}. This scattering boosts the UV photons to X-ray energies via inverse Compton scattering. Some X-ray photons reach our telescopes directly, and some photons irradiate the accretion disk and are reprocessed (e.g. via photoelectric absorption and fluorescence, thermalization, Compton scattering).  Reprocessed X-ray emission from the disk is generically referred to as a 'reflection' component in the X-ray spectrum.
 The most prominent reflection features are the iron~K$\alpha$ emission line at 6.4~keV and the Compton scattering `hump' peaking at $\sim 20$~keV (e.g., \citealt{risaliti13,marinucci14b,walton14,parker14} and see Fig.~\ref{fig:fek}-{\em left}.) We observe the iron~K emission line and other reflection features to be broadened due to Doppler motion in the accretion disk, and also by the gravitational redshift from the strong potential well of the black hole \citep{fabian89}. If the black hole is spinning rapidly and the ISCO is small, then a large gravitational redshift is observed. For the iron~K$\alpha$ line, a rapidly rotating black hole results in a strong red wing of the line that can extend down to 3--4~keV in the most extreme cases. This simple idea has been used to measure the spins of $\sim 50$ nearby AGN \citep{reynolds19}.

While measuring spin from broadened reflection features is a simple idea, astronomers have been plagued by systematic uncertainties that have cast doubt on black hole spin parameter estimations. For instance, from the spectrum alone, it is not always clear what emission is from the primary coronal continuum and what is relativistically broadened reflection. Moreover, we do not have strong observational constraints on the geometry of the X-ray corona, which effects the emissivity (irradiation profile) of the accretion disk and therefore can affect one's ability to uniquely identify the inner edge of the accretion disk (e.g. \citealt{dauser14}, \citealt{wilkins11}). 

Fortunately, in recent years we have made a breakthrough in our understanding of the X-ray emission from close to black holes with the discovery of X-ray reverberation. Implicit in the model described above is the idea that there should be a light travel time delay between the primary emission from the corona and the reflected emission off of the accretion disk. Reverberation time delays were first discovered in 2009 using the \xmm\ Observatory (\citealt{fabian09}, \citealt{zoghbi10} and \citealt{uttley14} for an early review). These short light travel lags generally confirm the picture that the X-rays are reprocessing off the inner accretion disk within a few gravitational radii. The compact size of the X-ray emitting region (e.g. \citealt{reis13}) is independently confirmed through microlensing \citep[e.g.,][]{chartas09,chartas12} and X-ray eclipses (e.g. \citealt{gallo21}).

 \begin{figure}
\centering
\includegraphics[width=\textwidth]{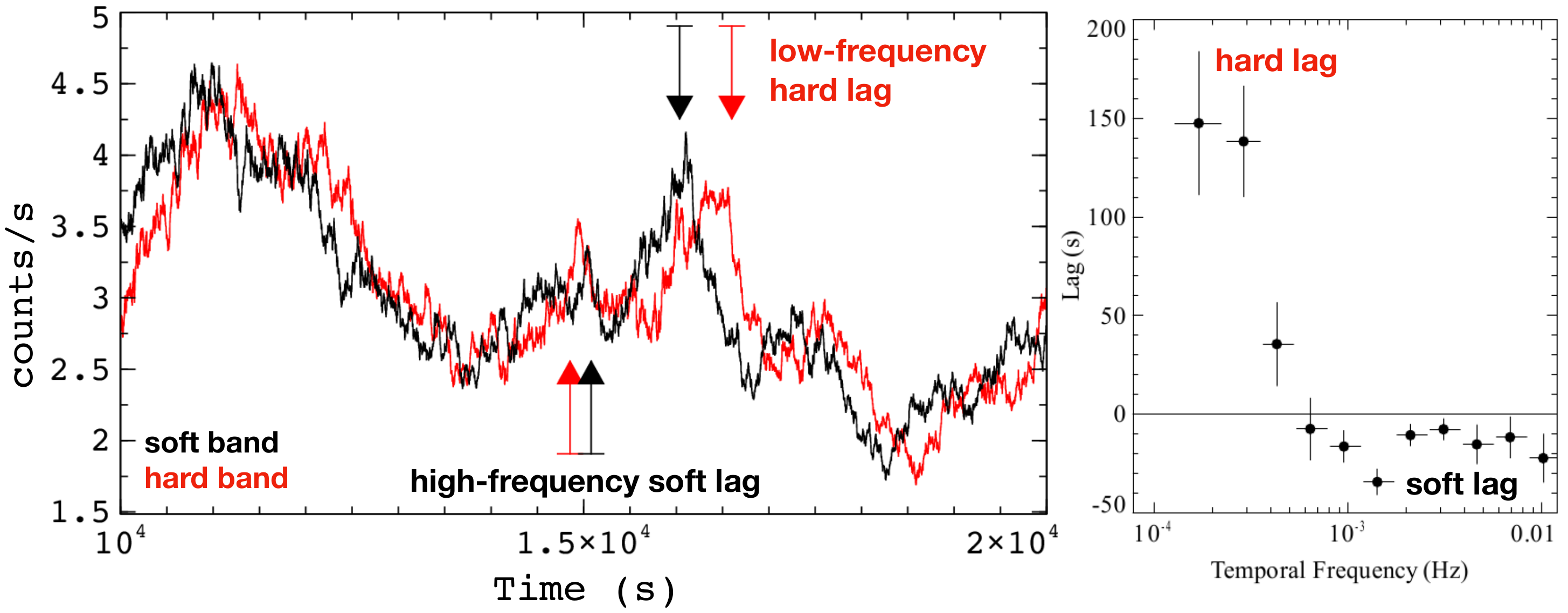}
\caption{{\em Left:} A portion of a simulated (noiseless) hard band (1-4 keV; red) and soft band (0.3--1~keV; black) light curve based on the properties of 1H0707-495, displaying variability on a range of timescales. On long timescales (low temporal frequencies), the hard band lag the soft, but on short timescales (high temporal frequencies), the soft band lags the soft. This demonstrates the benefit of studying X-ray reverberation (soft lags) using frequency-resolved timing analysis. {\em Right:} The observed lag-frequency spectrum of Seyfert galaxy 1H0707-495 \citep{zoghbi10}, showing the hard lags (positive lags) at low frequencies and the soft lags (nevative lags) at high frequencies. Figure courtesy of Abdu Zoghbi.} 
\label{fig:lag-freq}
\end{figure}

 \begin{figure}
\centering
\includegraphics[width=\textwidth]{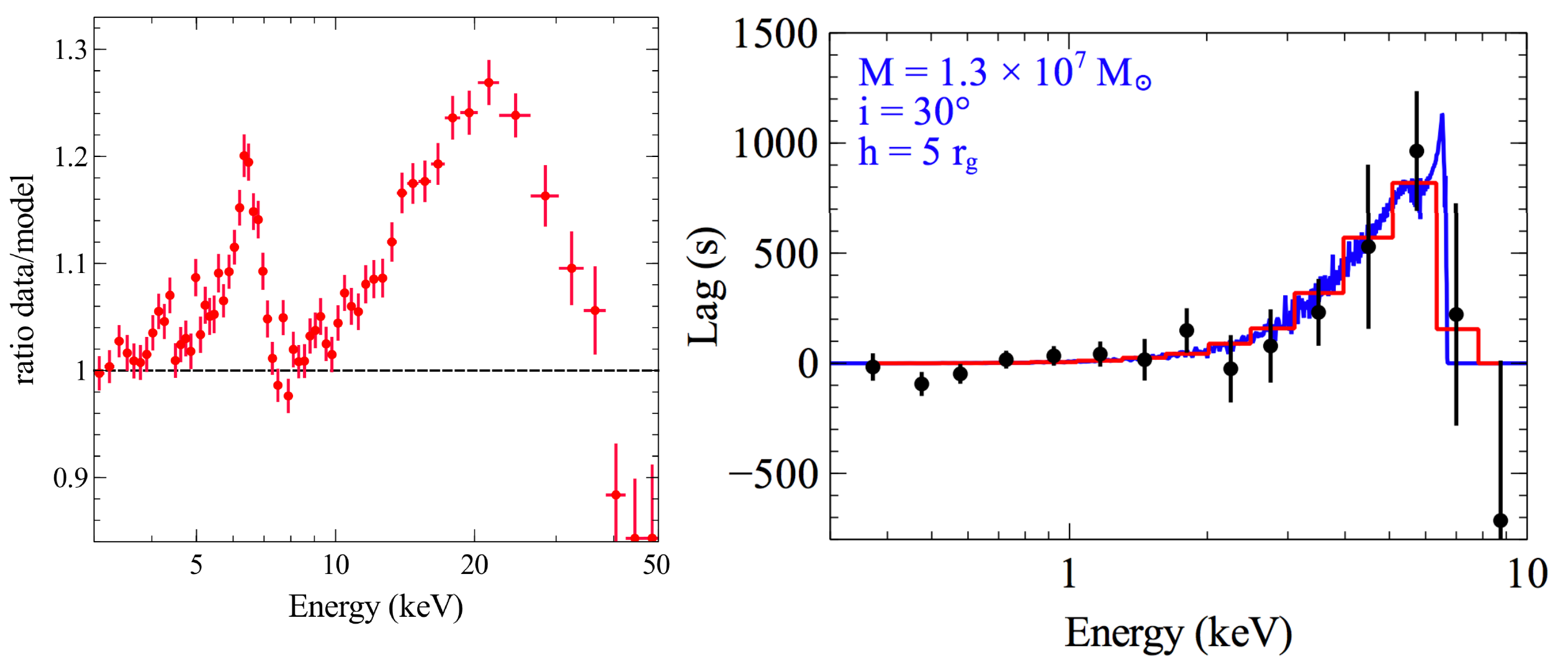}
\caption{{\em Left:} the \nustar\ spectrum of SWIFT~J2127.4+5654, with a power-law model of the continuum divided out. The ratio plot shows the broad Fe~K line and Compton hump \citep[adapted from][]{marinucci14b}. {\em Right:} The observed high-frequency time lags from PG~1244+026 \citep{kara14a} fitted with an iron~K reverberation model generated from a general relativistic ray-tracing simulation of a lamppost corona irradiating a razor thin disk \citep{cackett14}.} 
\label{fig:fek}
\end{figure}

X-ray variability from black holes is generally characterized as a red noise process, with variability from different processes contributing to the observed variability. It is for this reason that the technique used in the X-ray band is to measure Fourier-frequency-resolved time delays \citep[see above for a brief overview and][and references therein for a significantly more detailed description of the methods]{uttley14}. For the discovery object, the Narrow-line Seyfert 1 AGN 1H0707-495, \citet{fabian09} searched for time delays originally between the light curves in two energy bands, one dominated by the coronal continuum, and one dominated by the soft X-ray band, where there was evidence for relativistic reflection, most prominently in the form of a strong iron~L emission line. The frequency-resolved approach allowed for zeroing in on just the shortest timescale variability, where the reflection dominated band was observed to lag behind the continuum by roughly 30~seconds, or $2~R_{\mathrm{g}}/c$ for a $10^6~M_{\odot}$ black hole (see Fig.~\ref{fig:lag-freq}). In the years following this discovery, high-frequency reverberation lags were observed both in low-energy X-rays (e.g. \citealt{zoghbi11,cackett13,alston14,demarco13}), but also in the iron~K emission line (Fig.~\ref{fig:fek}-{\em right} and see \citealt{zoghbi12,kara13,kara16_global,vincentelli20}) and more tentatively, the Compton reflection hump above 10 keV (\citealt{zoghbi14,kara15}; thanks to the launch of the hard X-ray \nustar\ Observatory, and the development of techniques to deal with gaps in light curves; \citealt{zoghbi13b}).  However, follow-up hard X-ray studies with much longer exposures suggest little statistical deviations from log-linear lags \citep{zoghbi21}. We point readers to \citet{uttley14} for a review of many of these early observational discoveries.

\subsection{Using Reverberation to probe coronal geometry}
\label{sec:corona}

\begin{figure}
\centering
\includegraphics[width=\textwidth]{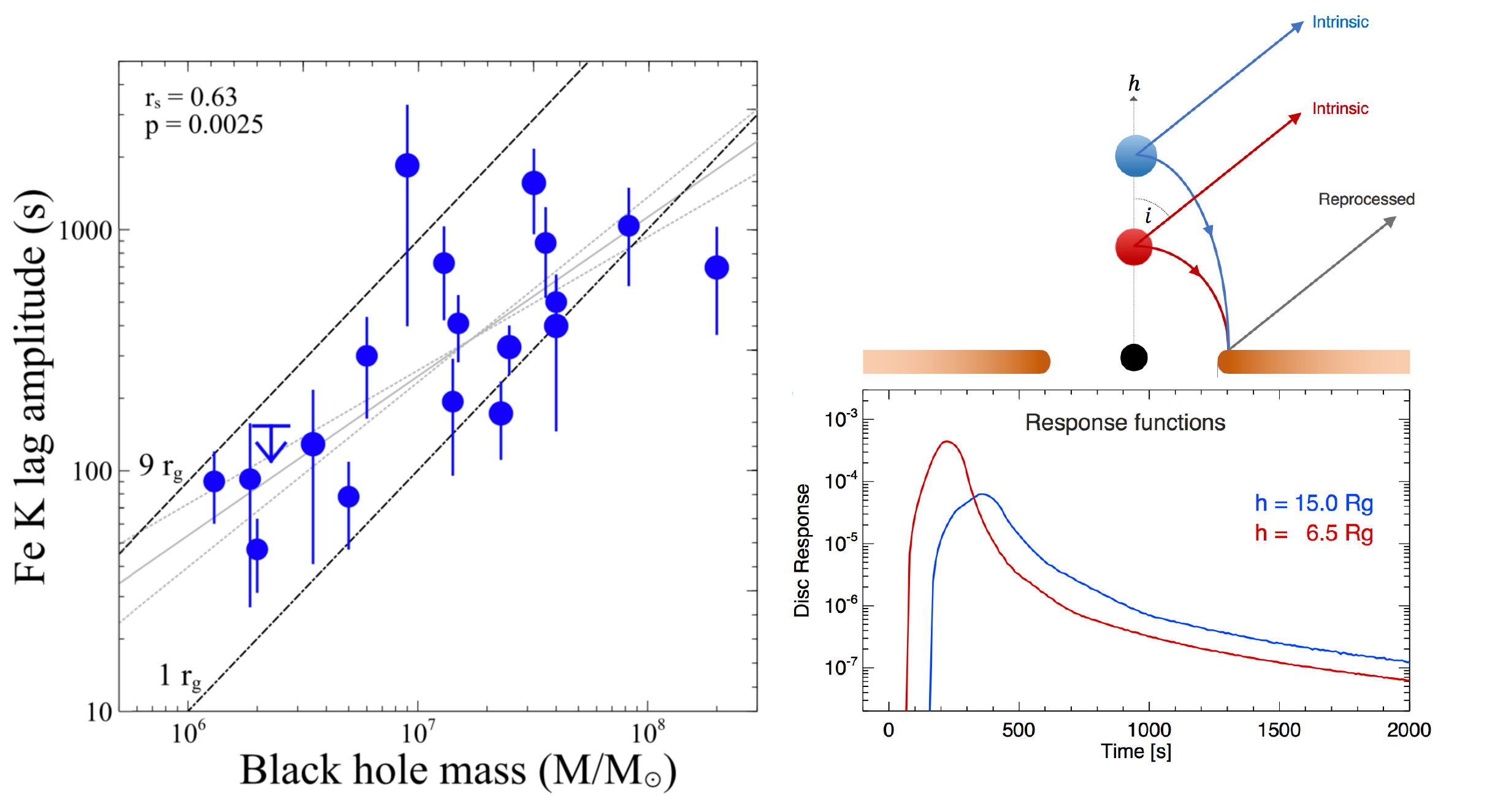}
\caption{{\em Left:} The iron~K lag amplitude vs. black hole mass for a sample of Seyfert AGN (updated from \citet{kara16_global}). The spread is likely in part due to uncertainties in black hole mass (not plotted), and also due to intrinsic differences in corona/disk geometry between different AGN {\em Right:} \citet{alston20} showed that reverberation can measure variable corona/disk geometries in a single AGN. The top panel shows a simple schematic of the geometry of the model, and the bottom panel shows the corresponding impulse response functions for the two different height coronae. }
\label{fig:xray_mass_corona}
\end{figure}

 To date, X-ray reverberation lags have been measured in roughly 2~dozen AGN \citep{demarco13,kara16_global}. In these sources, the lag amplitudes scale with the masses of the black holes, with some significant scatter in the relation (Fig.~\ref{fig:xray_mass_corona}-{\em left}). The spread is likely due to in part to uncertainties in the black hole mass, but also likely due to intrinsic variance in the geometry of the corona or inner accretion disk. Such effects are, for instance, important in IRAS~13224-3809 (\citealt{kara13b}, and confirmed with a much more extensive dataset in \citealt{alston20}). In this highly variable AGN, as the intrinsic luminosity of the corona increases, the reverberation lags become longer (Fig.~\ref{fig:xray_mass_corona}-{\em right}). By modeling the reverberation lags with General Relativistic ray-tracing simulations of a compact corona irradiating a thin accretion disk (see Section~\ref{sec:model}), \citet{alston20} showed that the height of the corona increases with increasing luminosity. This finding illustrates the ability of X-ray reverberation not just to make a picture of the black hole inner accretion flow, but a movie. Moreover, this work demonstrates that if the geometry of the inner accretion flow can be well constrained, X-ray reverberation lags can place constraints on fundamental black hole parameters, namely mass and spin.

\subsection{Modeling X-ray Reverberation with General Relativistic Ray-Tracing Simulations}
\label{sec:model}

X-ray reverberation near the black hole requires treatment of general relativistic effects. The basic approach is to model the response of the accretion disk (including relativistic dynamical distortions) to a sudden impulse of photons from the X-ray corona (usually assumed, for computational reasons, to be a point source above the accretion disk, positioned on the spin axis of the black hole; \citealt{campana95,reynolds99, wilkins13}). The first attempt to compare these impulse response functions to observations of lag-frequency spectra was presented in \citet{wilkins13}, and later \citet{cackett14} fitted the iron~K lag with a grid of impulse response functions in order to put constraints on, e.g. disk inclination and coronal height (Fig.~\ref{fig:fek}-{\em right}). This was extended to a larger sample fitting the lag-frequency spectra in \citet{emm14}. The next step was to calculate the energy-dependent lags across the entire reflection spectrum \citep{chainakun15,caballero18}. This model was used to fit the lag-frequency spectra in IRAS~13224-3809, discussed in Section~\ref{sec:corona} \citep{alston20}. 

In addition to general relativistic corrections, the observed X-ray lags are complex because they not only show reverberation lags, but also contain lags associated with the continuum variability (referred to here as `continuum lags'). These lags dominate on long timescales, and are thought to be associated with the propagation of mass accretion rate fluctuations through the accretion flow \citep{lyubarskii97,arevalo06}. 
Recently, \citet{ingram19} developed a general relativistic ray-tracing model that also includes the continuum lag. In this model, the energy spectrum from the coronal continuum is modelled as a powerlaw that fluctuates in photon index and normalization at a few per cent level, which causes time lags between soft and hard continuum photons. This can model both the long timescale continuum lags and high frequency reverberation signature \citep{mastroserio18,wang21}.

Beyond a simple lamppost corona irradiating a razor thin accretion disk, there have been recent efforts to account for more complex geometries of the corona and disk, and indeed, the data seem to suggest that the time lags would be better modelled with a corona extended over tens of gravitational radii  \citep{mastroserio20,wang21,zoghbi21}. \citet{wilkins16} calculated the impulse response functions for an extended corona with various geometries irradiating a razor thin disk, also considering the complicating factor of the propagation of the signal through the extended corona.  \citet{taylor18} studied the impulse response functions assuming a lamppost corona irradiating a more physical, nonzero scale height disk, following a Novikov-Thorne disk profile. More recently \citet{thomsen19} calculated the anticipated iron line and reverberation signature off the electron scattering photosphere of a General Relativistic Radiation Magnetohydrodynamic simulation of a super-Eddington accretion flow. These are important improvements to the models and are vital for understanding the lags from more realistic environments, but are computationally expensive and have not yet been fitted to data. The first attempt to fit a `more physical' extended corona geometry to observed X-ray time lags was presented in \citet{chainakun17}, starting with a `two-blob' corona geometry from two point sources at different heights above the accretion disk.  Modeling efforts continue to improve, with more physically-motivated assumptions, including low-frequency continuum lags modelled as the propagation of mass accretion rate fluctuations through the accretion flow \citep{wilkins16,mahmoud18}.

\subsection{X-ray lags from larger scales}

X-rays are not only a probe of the inner accretion flow, but also the hard X-ray continuum is absorbed, scattered, and reprocessed off circumnuclear gas flows beyond the inner disk, including from disk winds, the broad line region and dusty torus. This has led to an alternative explanation for X-ray lags as due to scattering of X-rays passing through an absorbing medium \citep[e.g.,][]{miller10b}, for instance a disk wind \citep{mizumoto19}.  But, these models require that most of the X-ray emitting gas is along our line of site \citep{zoghbi11} and do not explain the difference in lag-energy properties at high and low frequency \citep{kara13c}. However, there is mounting evidence that absorption can add to the complexity in observed lags, at least in some systems.  Time-dependent changes in  the absorber properties can lead to changes in observed lags \citep{silva16}, which has been suggested to explain lags in several objects which do not simply look like disk reverberation and which show absorption in their spectra \citep{kara15,zoghbi19}.

The most prominent feature from distant reflection is the narrow iron~K emission line, that is commonly seen in AGN spectra. This line has typically been associated with the torus, though recent {\it Chandra}/HETG studies of the line profile suggest Doppler broadening effects that would suggest that the lines originate in the outer accretion disk, or inner broad line region \citep{miller18}. This was supported by the correlation of the narrow iron line and continuum flux on short timescales \citep{zoghbi19}, with limits on the narrow iron~K lag suggesting it originates in a region within the optical BLR. Future telescopes with X-ray micro-calorimeters, like {\it XRISM} and {\it Athena} will put important spectral constraints that can confirm these results, and allow for measurement of X-ray reverberation lags to probe larger-scale gas flows around the AGN, complimenting longer wavelength reverberation that probes the outer disk and broad line region.

\subsection{X-ray reverberation beyond AGN}

While this review focuses on AGN reverberation, we would be remiss not to mention an important development in X-ray reverberation studies in the last few years, the discovery of X-ray reverberation in black hole X-ray binaries (BHXRBs). If black holes accretion flows are generally mass invariant, then BHXRBs are arguably the best way to study accretion, as they evolve on timescales that are millions to billions of times faster than their AGN counterparts. So, instead of studying different AGN populations in order to understand how the disk/jets/corona change with mass accretion rate, we can instead watch these states in a single BHXRB outburst over a few months. Reverberation lags in BHXRBs were first found using \xmm\ observations of the low mass BHXRB GX~339-4, where \citet{uttley14} discovered a $\sim 10^{-3}$~s lag between the coronal emission and reprocessed (e.g. fluorescence and thermalized) emission off of the accretion disk. This work was followed up by \citet{demarco15,demarco17}, where the reverberation lags were shown to shorten as the source reached peak luminosity in the hard state. A shortening lag suggests a smaller emitting region, and this was interpreted either as due to changes in the truncation radius of the accretion disk, or the height of the corona.

In 2017, we had a breakthrough in soft X-ray timing thanks to the launch of the {\em Neutron Star Interior Composition Explorer} ({\em NICER}; \citealt{gendreau16}). \nicer\ provides an unprecedented soft X-ray effective area and time resolution, virtually no pile-up, all while maintaining moderate (CCD-quality) energy resolution. All of these instrumental improvements have made for unprecedented measurements of X-ray reverberation in BHXRBs. Moreover, in March 2018 a new BHXRB, MAXI~J1820+070 went into outburst, and became the second brightest X-ray source in the sky (after only Sco~X-1).  \nicer\ observations of the luminous hard state of MAXI~J1820+070 showed the timescale of the reverberation lags shortened by an order of magnitude over a period of weeks, while the shape of the broadened iron K emission line remained remarkably constant, suggesting that the change in lags before the state transition were due to a decrease in the spatial extent of the corona, rather than a change in the inner edge of the accretion disk \citep{kara19}. This decrease in spatial extent of the corona tracks the decrease in radio luminosity, supporting earlier theoretical and observational work suggesting the corona is related to the base of the relativistic jet \citep{markoff05, fender1999quenching,homan2005multiwavelength}.

\section{UV/optical continuum reverberation}  \label{sec:disk}

\begin{figure}
\centering
\includegraphics[width=0.9\textwidth]{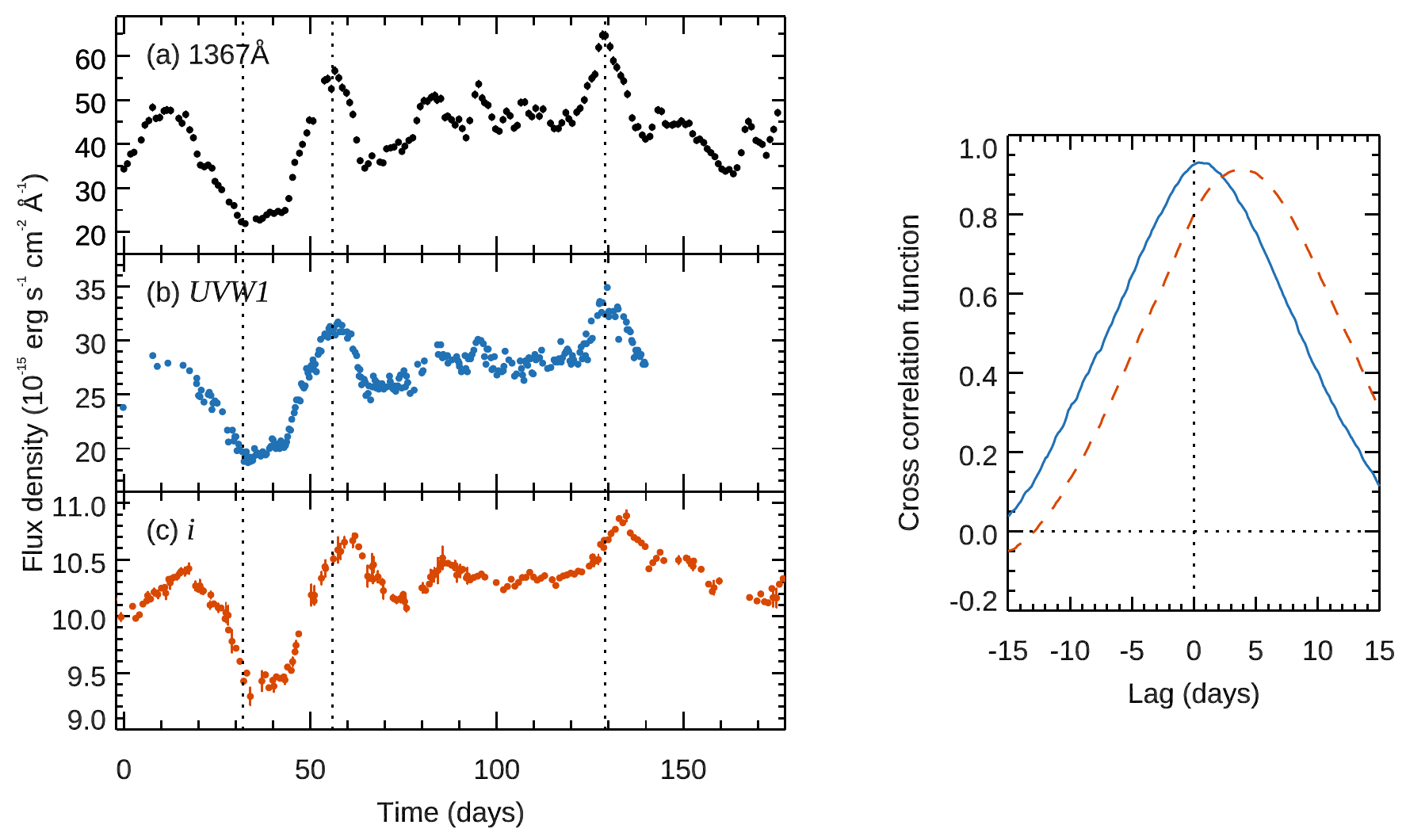}
\caption{Three of the continuum light curves of NGC~5548 from the AGN STORM campaign. The light curves are  (a) 1367\AA\  from \hst, (b) {\it UVW1} (2600\AA) from {\it Swift} and, (c) $i$ (7648\AA) from ground-based telescopes.  The light curves are all highly correlated.  Vertical dotted lines are plotted to guide the eye.  The right-hand panel shows the cross correlation function with respect to the 1367\AA\ light curve for the {\it UVW1} (blue, solid line) and $i$ (orange, dashed line) bands, with the correlation peaking at a lag of around 0.9 days for {\it UVW1} and 4 days for $i$.  Data from \citet{fausnaugh16}.}
\label{fig:ngc5548_cont}
\end{figure}

UV/optical continuum reverberation mapping looks for lags between continuum light curves measured at different wavelengths.  The accretion disk reprocessing scenario posits that a central source (presumably emitting in the X-rays or far-UV) irradiates the accretion disk and drives the UV and optical variability, with the hot inner part of the accretion disk seeing the irradiating light before the cool outer part of the disk \citep[see e.g.,][for detailed descriptions]{collier99,cackett07}.  This therefore predicts that the UV and optical continuum light curves should be well-correlated, with the UV leading the optical (as is observed, see Fig.~\ref{fig:ngc5548_cont}), and the lag dependent on the temperature profile of the disk.  For a standard optically thick and geometrically thin \citep{SS73} accretion disk, the temperature profile follows $T(R) \propto (M \dot{M})^{1/4} R^{-3/4}$, where $\dot{M}$ is the mass accretion rate.  Since the lag gives a measure of the radius ($\tau = R/c$), and for blackbody radiation we have $\lambda \propto T^{-1}$, this $T(R)$ corresponds to $\tau (\lambda) \propto  (M\dot{M})^{1/3} \lambda^{4/3} \propto  (M^2\dot{M}_{\rm E})^{1/3} \lambda^{4/3}$, where $\dot{M}_{\rm E} = L_{\rm bol}/L_{\rm Edd}$ is the Eddington fraction.  One can generalize this further, as a disk with $T(R) \propto R^{-b}$ leading to wavelength-dependent lags of $\tau(\lambda) \propto \lambda^{1/b}$.

Early BLR reverberation campaigns showed that the UV and optical continua were well-correlated with inter-band lags of less than a few days \citep{stirpe94,wanders97}. Observed lags showed evidence for the expected wavelength-dependence, though at relatively low significance \citep{collier98,collier01}.  Multi-year optical photometric monitoring of 14 AGNs by \citet{sergeev05} showed both that lags increase with wavelength and that more luminous objects exhibited longer lags, as expected in the disk reprocessing scenario. Again, however, few of the lags were significant at $>$3$\sigma$.  Observationally, it is  a challenge to measure these lags with traditionally scheduled ground-based telescopes, since temporal sampling of $<24$\,hours is required over a baseline of weeks to months. 

The field has changed in the last decade for two reasons.  Firstly,  {\it The Neil Gehrels Swift Observatory} \citep[hereafter {\it Swift};][]{burrows05} has made it possible to perform high cadence (often multiple observations per day), high signal-to-noise observations in the X-ray, UV and optical.   Secondly, the introduction of robotic observatories dotted across the globe has made it easier to perform high cadence optical monitoring with significantly fewer losses to poor weather, while also avoiding many of the logistical issues involved in running long-term monitoring campaigns.

\subsection{Intensive disk reverberation mapping with {\it Swift}}

\begin{figure}
\centering
\includegraphics[width=\textwidth]{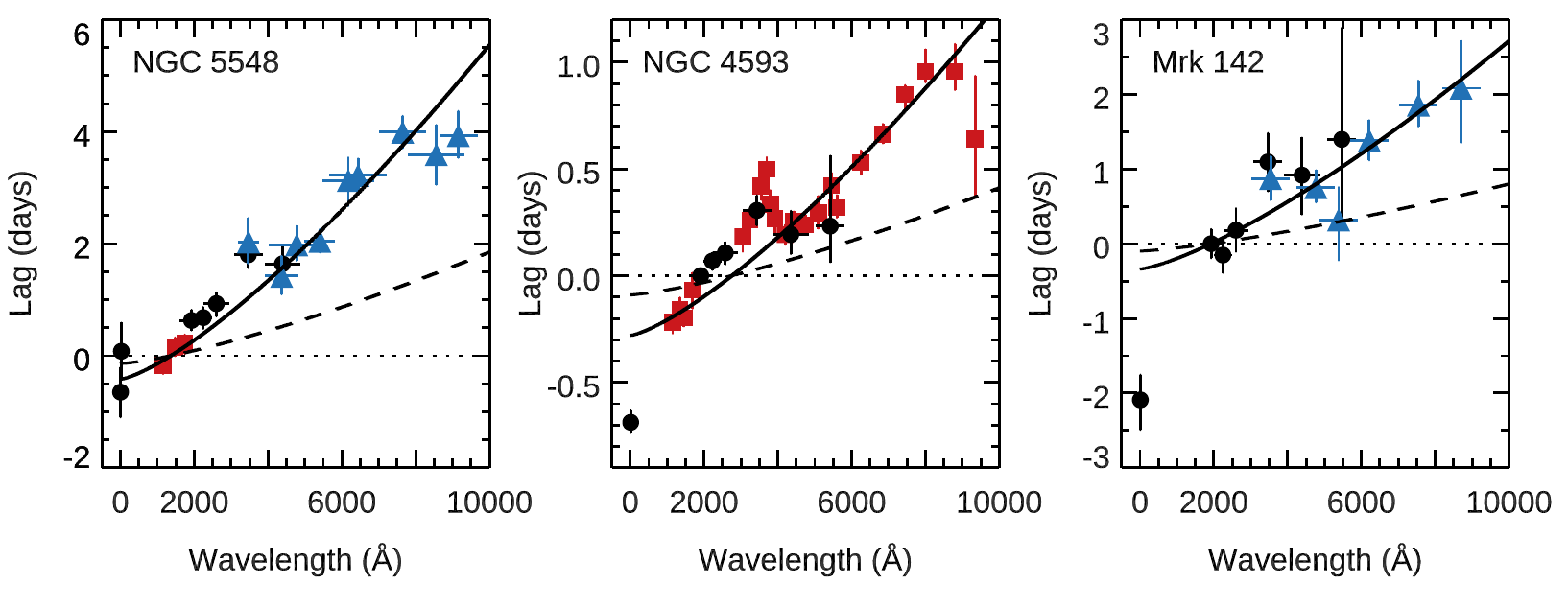}
\caption{Wavelength-dependent continuum lags in NGC 5548 \citep{fausnaugh16}, NGC 4593 \citep{cackett18}, and Mrk 142 \citep{cackett20}. Data from {\it Swift} (black circles), \hst\ (red squares) and ground-based observatories (blue triangles).  The solid lines show the best-fitting $\tau \propto \lambda^{4/3}$ relation (excluding the X-rays and $u/U$ bands), while the dashed lines show the predicted lags based on reasonable estimates of the black hole mass and accretion rate \citep[eqn.~12 in][]{fausnaugh16}). The difference between the observed (solid line) and predicted (dashed line) relations is the `accretion disk size problem'.  The lags in the $u/U$ bands (around 3500\AA) show a clear excess, and in NGC~4593 the lag spectrum reveals a broad excess around the Balmer jump.  Finally, note also how there is no consistent relationship between the X-rays and the best-fitting $\lambda^{4/3}$ relation. }
\label{fig:contlags}
\end{figure}

{\it Swift} can rapidly slew to targets, which allows it to observe many targets per day, each with relatively short (1~ks) visits. This is perfect for continuum reverberation mapping since {\it Swift} can even observe the same target multiple times per day. Moreover, it has both an X-ray and a UV/optical telescope (UVOT), allowing for X-ray spectroscopy and UV/optical photometry in 6 filters from 1928\AA\ ({\it UVW2}) to 5468\AA\ ($V$). The first uses of {\it Swift} for continuum reverberation demonstrated its potential to measure lags to much higher precision than had previously been achieved, and also demonstrated that the lags were longer than expected from the standard thin disk model \citep{shappee14,mchardy14}.  A breakthrough was achieved with monitoring of NGC~5548 with {\it Swift} along with {\it Hubble Space Telescope} (\hst) and ground-based observations as part of the Space Telescope and Optical Reverberation Mapping program (AGN STORM, \citealt{derosa15}).  That campaign more than doubled the number of visits with all 6 UVOT filters in either of the previous two campaigns, achieving a sampling rate of better than once per 0.5 day \citep{edelson15,fausnaugh16}.  The main results of that campaign -- that the wavelength-dependent lags approximately follow $\lambda^{4/3}$, that the accretion disk is bigger than expected by a factor if $\sim$3, that the lag in the $U$ band (3465\AA) is in excess of the $\lambda^{4/3}$ relationship  \citep{edelson15,fausnaugh16},  and that the X-ray light curve is not consistent with being the driving light curve \citep{gardnerdone17, starkey17} -- are common to most of the {\it Swift} continuum reverberation campaigns that have followed \citep{edelson17,edelson19,cackett18,cackett20,mchardy18,hernandezsantisteban20}.  Fig.~\ref{fig:contlags} shows some examples of  these recent results.

\subsection{Continuum reverberation with large surveys}

 A different approach to the intensive high-cadence monitoring of small numbers of objects is to look at a larger sample of objects but at a lower cadence. This approach has been undertaken utilizing large surveys.   While generally individual objects will not yield high-fidelity lag-wavelength relations, the population of objects can be studied as a whole.   This powerful approach is being taken by PanSTARRS \citep{jiang17}, DES \citep{mudd18,yu_DES_20} and SDSS \citep{homayouni18}. Looking at 240 quasars observed by PanSTARRS, \citet{jiang17} finds that the average lags are also a factor of 2 -- 3 larger than expected.  Similarly, \citet{mudd18} find disk sizes consistent with being a factor of 3 too large.  On the other hand, \citet{homayouni18} and \citet{yu_DES_20} report disk sizes consistent with the expectations from a standard disk. \citet{homayouni18} demonstrate that if only the best lag measurements are used, then disk sizes are biased to be larger than when the full sample is considered.  They are also able to show that disk sizes scale with black hole mass, as expected.  As we look to the future, upcoming large surveys such as the Legacy Survey of Space and Time at the Vera C. Rubin Observatory will significantly expand the number of AGNs with coarsely sampled multi-color photometric light curves for exploring continuum reverberation. 

\subsection{The accretion disk size problem}

The observed wavelength-dependent lags are usually fit with a simple model of the form $\tau = \tau_0 \left[(\lambda/\lambda_0)^\beta - 1 \right]$, where $\lambda_0$ is the reference band wavelength (usually chosen to be the best-sampled and highest S/N band), $\tau_0$ is the normalization, and the value of $\beta$ is 4/3 for a standard thin disk.  Generally it is found that $\beta \sim 4/3$, though the best-fitting results from NGC 5548 prefer a slightly shallower value close to 1 \citep{fausnaugh16}.  $\tau_0$ can be predicted based on the mass and mass accretion rate of the target \citep[see equation 12 in][]{fausnaugh16}. When the predicted  $\tau_0$ is compared to the best-fitting value, it has frequently been found to be 2 -- 3 times larger than expected.  This has been seen both in the intensive disk reverberation studies \citep{shappee14,mchardy14,edelson15,edelson17,edelson19,fausnaugh16,fausnaugh18_contm,cackett18,cackett20,pozonunez19} and also from surveys \citep{jiang17,mudd18}, but is not always the case \citep{edelson19,hernandezsantisteban20,homayouni18,yu_DES_20}.  Lags that are longer than expected indicate that the radius for a given temperature is larger than expected, or that some of the assumptions are wrong, such as the mass accretion rate estimate or the choice of factor to convert from wavelength to temperature.  This is often referred to as the accretion disk size problem (see the comparison of predicted and measured relations in Fig.~\ref{fig:contlags}). Interestingly, independent results using gravitational microlensing of quasars indicates that the optical emitting region in those objects is similarly larger than expected \citep[e.g.][]{morgan10}.

Several explanations for the disk size problem involve disks being more complex than the simple \citet{SS73} picture.  For instance, to explain gravitational microlensing results \citet{dexter11} suggest that local fluctuations lead to an inhomogeneous, time-dependent disk structure and can reproduce the larger than expected disk sizes.  Alternatively, the assumption that the disk emits as a blackbody may also be incorrect. \citet{hall18} suggest that the larger disk sizes seen by continuum reverberation can be explained by accretion disks with a sufficiently low-density scattering atmosphere, which leads to a substantially different temperature profile.  Such a scattering atmosphere gives longer lags, a flatter wavelength dependence, and can match the observed NGC~5548 lags. Alternative disk models also predict different temperature profiles \citep{mummery20}.   More sophisticated disk reprocessing models using general relativistic ray-tracing give systematically longer lags than the simple analytic relation frequently used, and are able to reproduce the lags with reasonable mass and mass accretion rate estimates, potentially solving the accretion disk size problem \citep{kammoun19,kammoun21a,kammoun21b}. Interestingly, the implied height of the X-ray source is typically larger than seen from X-ray reverberation, which may suggest that the corona is vertically extended (see Section~\ref{sec:corona}). 

Different geometries have also been considered, for instance, \citet{gardnerdone17} suggest the inner disk puffs up to prevent hard X-rays directly irradiating the UV/optical disk and propose that the lags may be associated with the timescale for the outer disk vertical structure to vary.   Another class of models explains AGN UV/optical variability and disk lags as due to a different process (not the light-crossing time).  In the corona-heated accretion-disk reprocessing model \citep{sun20a,sun20b} the X-ray-emitting corona and the accretion disk are coupled via the magnetic field.  Magnetic turbulence in the corona causes X-ray variability and temperature fluctuations in the disk, with the temperature fluctuations delayed with respect to the X-ray variability as the magnetohydrodynamic waves propagate outwards at the Alfv\'{e}n velocity.  This produces longer lags than the simple disk reverberation model.

Other considerations are lags not associated with the disk. For instance, the addition of winds can lengthen the lags \citep{sun19}.  The contributions from broad emission lines that fall within photometric filter bands are not enough to significantly skew the lags \citep{fausnaugh16}, however, as we discuss in detail next, continuum emission from the BLR gas may have an important influence.

\subsection{The diffuse Balmer continuum}

\begin{figure}
\centering
\includegraphics[width=0.7\textwidth]{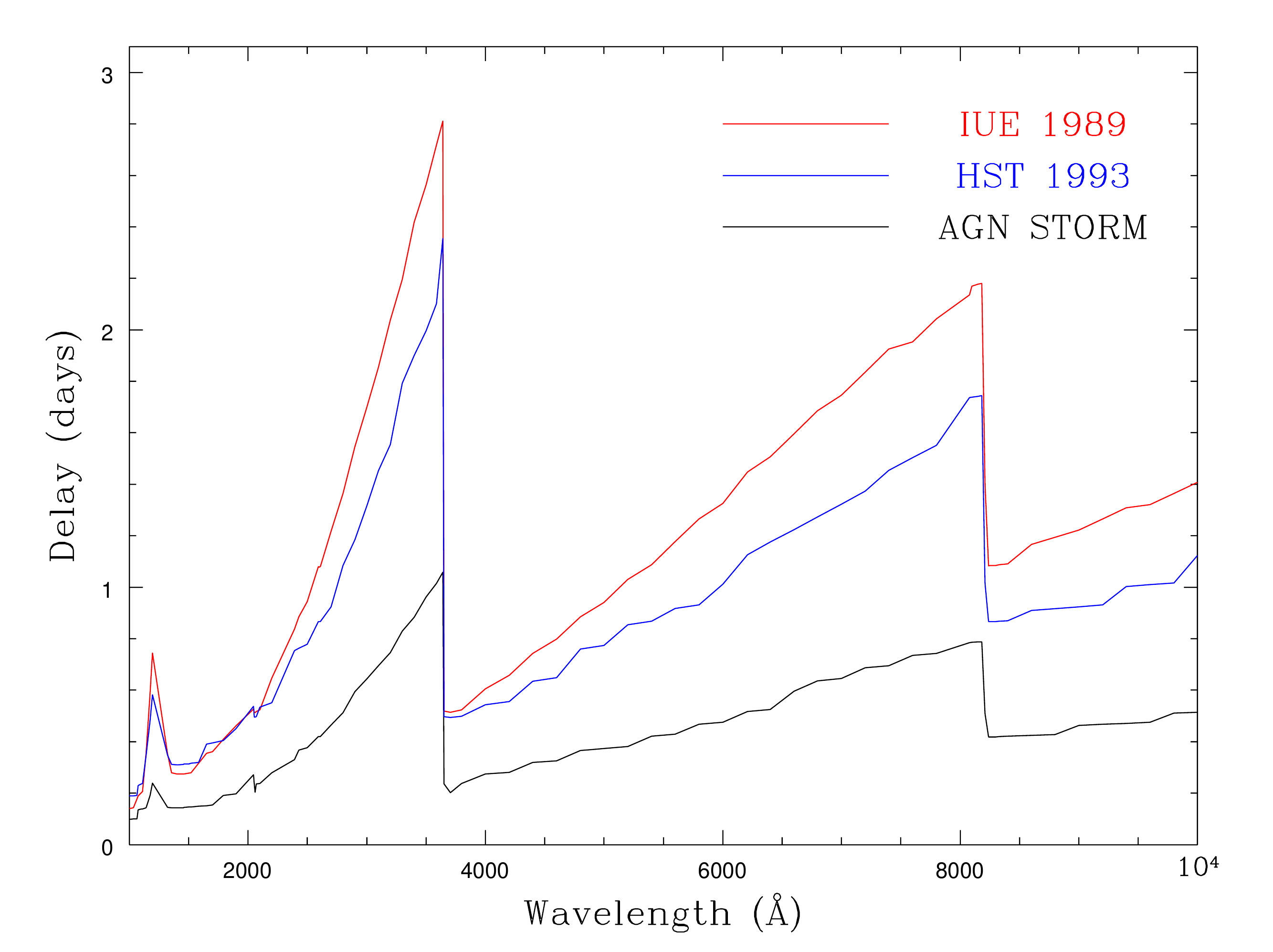}
\caption{Model wavelength-dependent lags from the BLR diffuse continuum measured relative to three different lag-less driving continuum light curves \citep[taken from][]{koristagoad19}. Note lags are expected at all wavelengths, but especially prominent around the Balmer and Paschen jumps.}
\label{fig:DClags}
\end{figure}

A common feature of continuum lags is that the $U$ band (3465\AA) lag is consistently in excess of an extrapolation of the trend through the rest of the UV/optical, by a factor of around 2 \citep[e.g.,][]{edelson19}.  This can be explained if emission from the `diffuse continuum' in the BLR is significant.  The `diffuse continuum' (thermal free-bound and free-free, plus scattering) emanates from the same clouds that emit the broad emission lines \citep{koristagoad01,koristagoad19,lawther18,netzer20}. It is expected to contribute substantially to the continuum at all wavelengths, but particularly around the Balmer (3646\AA) and Paschen jumps (8204\AA), as shown in Fig.~\ref{fig:DClags}.  Since the BLR is at a greater distance from the central irradiating source than the accretion disk, this diffuse continuum emission will act to lengthen the observed lags.  The clearest example of this comes from {\it Swift} and \hst\ monitoring of NGC~4593 -- the UV spectroscopic coverage allowed \citet{cackett18} to resolve the lag structure around the Balmer jump for the first time, rather than just seeing an excess in one broad photometric band (see Fig.~\ref{fig:contlags} for examples of the $U$ band excess and lags around the Balmer jump in NGC~4593).  Moreover, modeling of the light curves showed two components -- a rapid response (presumably from the disk), and a slower response from an extended reprocessor \citep{mchardy18}.  It has even been suggested that the lags are entirely due to non-disk emission. \citet{chelouche19} applied a multivariate lag analysis to Mrk~279 and suggest that lags originate from photoionized material above the outer disk, rather than from disk reprocessing closer in.

In addition to the signature of the diffuse BLR in the lags, this component should also show up prominently in the variable (rms) spectrum.  Recent studies \citep[e.g.][]{mchardy18, cackett20, hernandezsantisteban20} that perform a flux-flux analysis to calculate the variable spectrum show that the $u/U$ bands is also enhanced there (e.g. approximately 15\% in Mrk~142), but overall the variable spectrum follows a simple power-law. Future work that combines lag and spectral analysis together is needed to properly separate out the contributions from the disk and diffuse BLR.

\subsection{The weak X-ray/UV correlation}

The {\it Swift} continuum reverberation campaigns have also shown that the relationship between the X-ray and UV/optical variability is complex. The peak correlation coefficient between the X-ray and {\it UVW2} light curves is consistently much weaker than that between {\it UVW2} and the other UV/optical light curves \citep{edelson19}. Moreover, the X-ray lag is often seen to be offset with respect to the best-fitting UV/optical relation (see~Fig.~\ref{fig:contlags}). It can even be longer than expected from a straightforward lamp-post scenario \citep{noda16}. Even when the X-ray lag is consistent, analysis shows that in some cases, e.g. NGC~5548, the X-ray light curve does not look like the light curve needed to drive the variability at longer wavelengths \citep{starkey17,gardnerdone17}.  Other AGNs show no correlation between X-ray and UV/optical variability at all \citep{buisson18,morales19}.  This lack of a clear connection between the X-ray and UV/optical is puzzling and a problem for the simple disk reprocessing picture, and therefore may indicate more complex geometries or absorption \citep{gardnerdone17}, or that different variability processes are taking place on different timescales \citep{pahari20}. In the future, it will be important to connect the implied geometry from X-ray reverberation to the UV/optical continuum geometry.

On-going and upcoming projects involving continuum reverberation promise to further pursue these issues and shed light on the UV/optical continuum-emitting region in AGNs.

\section{Broad line region reverberation}  \label{sec:blr}

The first reverberation mapping experiments for AGNs  focused on trying to measure the BLR response to variations in the continuum flux. However, these early investigations were hampered by erroneously large expectations for the size of the BLR. While numerous studies of NGC\,4151 indicated that the BLR size in that AGN was smaller than 30 light days in radius \citep{cherepashchuk73,antonucci83,bochkarev84}, the low-luminosity and high ionization parameter of NGC\,4151 provided a plausible way out of the difficulties introduced by this small size. It was not until \citet{peterson85} carried out repeated observations of the luminous Seyfert Akn\,120 over a period of 4 years and clearly demonstrated that the predicted BLR size from photoionization equilibrium arguments was a factor of $\sim 10$ too large, that an unavoidable deficiency in our understanding was revealed.  Additional studies of other AGNs confirmed these results (e.g., \citealt{perez89,maoz90}). 

Thus it was not until the early 1990s that the first accurate measurements of broad-line reverberation were obtained.  The breakthrough was facilitated largely by the efforts of the International AGN Watch consortium, a large-scale coordinated endeavor involving simultaneous UV and optical photometric and spectroscopic monitoring, from space and from the ground\footnote{\url{http://www.astronomy.ohio-state.edu/~agnwatch/history.html}}.  The initial results, summarized by \citet{peterson93}, showed decisively that the broad lines respond rapidly to continuum variations, with high ionization lines like \ion{C}{IV} and \ion{He}{II} responding first and then lower ionization lines such as H$\beta$ responding later, indicating ionization stratification in the BLR (see Fig.~\ref{fig:n3783} for a recent example).   Additionally, gross radial motions that would indicate infall or outflow as the dominant motions in the BLR, were ruled out \citep{korista95}, demonstrating that rotation dominates the bulk motions of BLR gas in the local Seyferts that were targeted for study.  Around the same time, the shortcoming in early photoionization models was determined to be the use of a single representative cloud for the BLR, whereas a ``locally optimally emitting cloud (LOC)'' model  \citep{baldwin95} more successfully reproduced early reverberation results, including the BLR size and ionization stratification.  Unlike single-cloud models, the LOC model included a wide range of gas conditions  and found that BLR emission preferentially arises from the locations where the gas conditions are optimal. 

The key to observational success measuring lags involved satisfying the following constraints in the planning and execution of reverberation mapping programs:
\begin{itemize}
    \item a program length at least 3 times the longest expected time delay \citep{horne04};
    \item a temporal cadence fine enough to resolve the shortest expected time delay;
    \item high signal-to-noise ($\gtrsim 50$) in the continuum and significantly higher in the emission lines, so that variability amplitudes of a few percent could be clearly detected;
    \item flux calibration with a precision of 2\% or better across all the spectra obtained throughout the program \citep{peterson04};
    \item and a dash of good luck, since AGN variability is stochastic and unpredictable, though longer program lengths increased the chances of detecting variability (e.g., \citealt{macleod16}).
\end{itemize}

With a better understanding of the observational constraints necessary for a successful reverberation mapping program, the number of accurate broad line reverberation measurements began to rapidly accrue through the efforts of several groups.  This also led to the development of new tools for improving the time delay measurements.  Thus a consistent reanalysis of all previous  reverberation measurements for UV and optical broad lines was carried out by \citet{peterson04}, with the same tools and methods applied in all cases, so that results from different groups could be compared in a more straightforward manner.  The various analysis methods that had been employed by different groups were also examined and compared, leading to a set of best practices for the field and laying the groundwork for continued and improved success of BLR reverberation mapping experiments.  With a solid footing for broad-line lag measurements, RM goals have thus expanded to tackle new questions and focus on AGNs beyond the brightest local Seyferts, and the observational constraints needed to achieve these goals have continued to evolve as well.

\subsection{Black Hole Masses}

With the availability of accurate broad-line time delay measurements, it became clear that reverberation mapping could be used to constrain the mass of the central black hole in AGNs \citep{peterson99,peterson00}.  With the measurement of a time delay $\tau$ and line width $V$ for a broad emission line:
\begin{equation}
    M_{\rm BH} = f \frac{c\tau V^2}{G}
\end{equation}
\noindent where $c$ is the speed of light, $G$ is the gravitational constant, and $f$ is an order-unity scaling factor that accounts for the (generally unknown) geometry and kinematics of the broad line region gas.  Notably, when different emission lines are probed in the same AGN, the high ionization lines have short time delays and are found to be broader in width than the low ionization lines, generally agreeing with the expectations for virial motion and leading to consistent $M_{\rm BH}$ determinations (e.g., \citealt{peterson00,kollatschny01}).

\begin{figure}
\centering
\includegraphics[width=\textwidth]{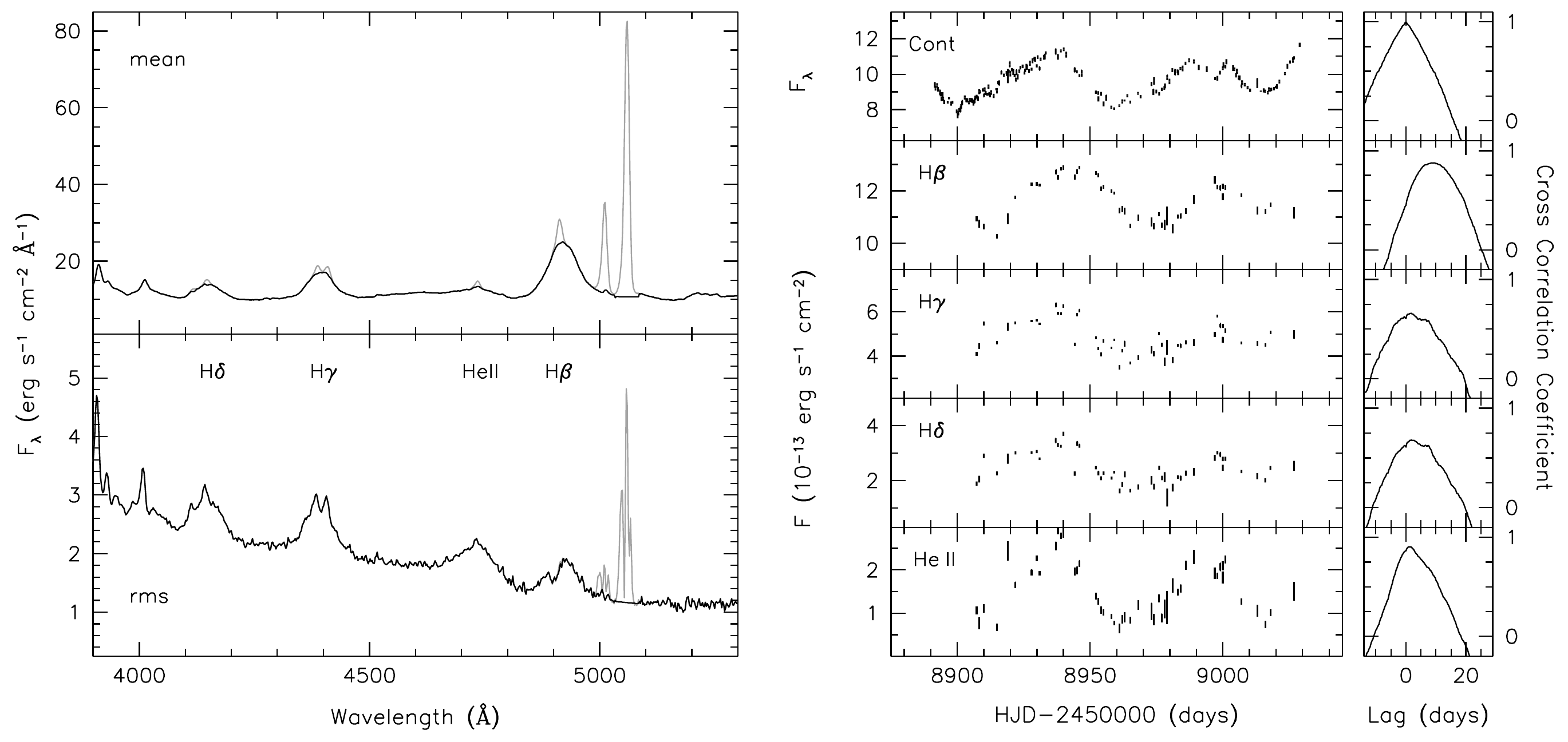}
    \caption{{\it Left:} Mean and root-mean-square (rms) optical spectra of NGC\,3783 collected over the course of 4 months in early 2020. The rms spectrum shows the  spectral components that were variable during the observing campaign.  {\it Right:} Light curves and cross-correlation functions for the continuum and the strong broad emission lines.  From \citet{bentz21}.   }
    \label{fig:n3783}
\end{figure}

As one of the recommended best practices, \citet{peterson04} demonstrated that it was critical to measure the line width in the {\it variable} part of the emission line (see Fig.~\ref{fig:n3783}), so that the line-of-sight velocity was directly related to the reverberating gas, when determining $M_{\rm BH}$.  Furthermore, focusing on the variable emission excludes narrow lines and other nonvariable emission that would otherwise bias the line width measurement, a point that is especially important for lines like \ion{C}{IV} that have multiple components (cf.\ \citealt{denney12}).

The population average value of $f$ has been constrained through comparison of the \msigma\ relationship \citep{gultekin09,kormendy13,mcconnell13} for AGNs and for quiescent galaxies.  Under the assumption that all the galaxies are drawn from the same parent sample, the multiplicative scaling factor that is needed to bring the intercept of the AGN relationship into agreement with that of the quiescent galaxies gives $\langle f \rangle$.  When the second moment of the emission line is used to represent the velocity width (another best practice recommended by \citealt{peterson04}), values of $\langle f \rangle$ have ranged from $2.8$ \citep{graham11} to $5.5$ \citep{onken04}, with most values settling around $4-5$ (e.g., \citealt{park12,grier13,batiste17}).  For the two AGNs that have black hole masses measured through stellar dynamical modeling, the reverberation masses based on $\langle f \rangle$ are generally consistent with the masses constrained by the dynamical models (NGC\,4151: \citealt{bentz06,onken14}, NGC\,3227: \citealt{davies06,denney10}).  

If inclination angle is assumed to be the largest contributor to the value of $f$, then $\langle f \rangle \approx 4-5$ implies that the average AGN in the reverberation sample is viewed at an inclination angle of $25-30^{\circ}$, which also seems reasonable based on our current understanding of the AGN unification model (e.g., \citealt{urry95}).  However, individual AGNs may have inclination angles that deviate from the population average, and so reverberation masses for individual objects that make use of $\langle f \rangle$ have an additional factor of $2-3$ uncertainty.

\subsection{\rl\ Relationship}

Another important product of reverberation mapping results is the scaling of the typical BLR size with the luminosity of the central AGN, the \rl\ relationship \citep{koratkar91,kaspi00,kaspi05}, which was expected based on photoionization arguments and searched for with the earliest reverberation measurements.  When the host-galaxy starlight contamination is properly removed from the AGN luminosity, the relationship for local Seyferts has the form $R_{\rm BLR} \propto L^{1/2}$  (\citealt{bentz06a,bentz09,bentz13}).  Microlensing of the BLR in lensed quasars, where the magnification amplitude is dependent on the size of the emission region, provides an independent way to probe the \rl\ relationship, and the results agree well  \citep{guerras13}.  Additionally, the Pa\,$\alpha$ BLR size in the quasar 3C\,273 was resolved in the near-infrared with the GRAVITY instrument on VLT, and the size compares well with those derived from reverberation mapping of the Balmer lines \citep{gravity_3c273}, although  reverberation mapping  determines  responsivity-weighted radii while interferometry instead measures flux-weighted radii.

The \rl\ relationship has been used to estimate $M_{\rm BH}$ for large samples of broad-lined AGNs with only a single spectrum per target (e.g., \citealt{shen11}).  Rather than investing months or years to measure the BLR size for an AGN of interest, the \rl\ relationship allows $R$ to be predicted from the AGN luminosity, which may then be combined with the broad line width and an adopted $f$ factor to estimate $M_{\rm BH}$ (e.g., \citealt{vestergaard06}).

The majority of BLR sizes thus far have been measured for the H$\beta$ emitting region, so at this time, the most well-constrained \rl\ relationship is for H$\beta$.  Efforts to bootstrap these results for the use of \ion{Mg}{II} and \ion{C}{IV} in the rest-frame ultraviolet (e.g., \citealt{vestergaard06,onken08,woo18}) have facilitated estimates of $M_{\rm BH}$ for high redshift quasars.  Thus, $M_{\rm BH}$ estimates for $z\gtrsim6$ quasars are currently dependent on the H$\beta$ \rl\ relationship for local broad-lined AGNs.  Recent investments in multiplexed reverberation mapping programs at high redshift will likely improve upon these estimates in the coming years as calibrated relationships for \ion{C}{IV} and other emission lines become available \citep{kaspi07,hoormann19,grier19}.  

\begin{figure}
\centering
\includegraphics[width=0.7\textwidth]{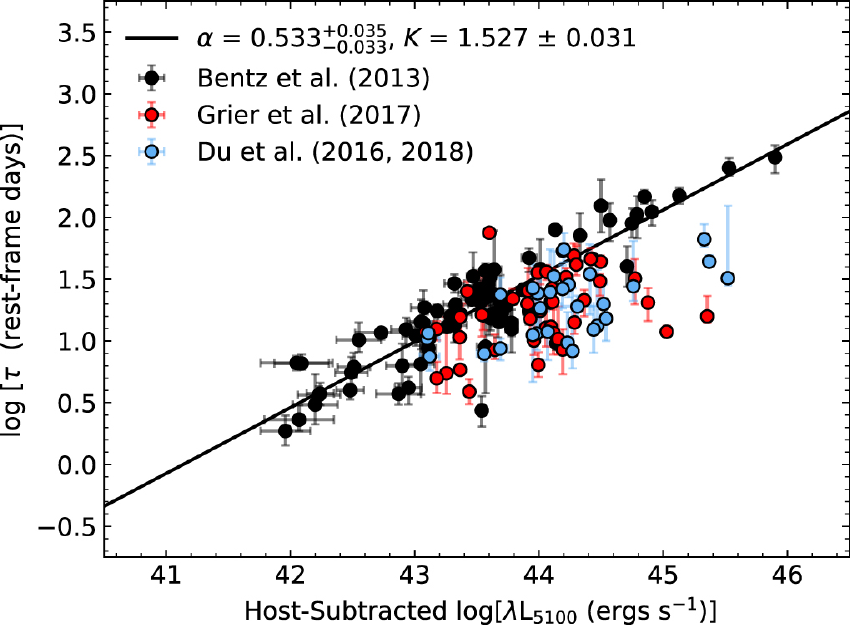}
\caption{The \rl\ relationship for all AGNs with H$\beta$ reverberation lags.  The black points show local Seyferts, while the blue point shows a sample of rapidly accreting AGNs and the red points show a sample of AGNs at higher redshift. From \citet{fonseca20}.}
\label{fig:rl}
\end{figure}

However, efforts to widen the population of AGNs that are studied by reverberation mapping have included objects expected to have higher mass accretion rates than those found among local Seyferts, as well as objects at higher redshift. Several studies show that many AGNs, including some of the highest accretion rate objects, fall below the usual \rl\ relation (see Fig.~\ref{fig:rl}; \citealt{du15,du16,du18,grier17_sdssrm,fonseca20}.)  While the reason is not currently understood,  some ascribe it to physical changes arising from high accretion rates (e.g., \citealt{dallabonta20}) while others have suggested that is related to differences in the shape of the ionizing continuum (e.g., \citealt{fonseca20}).  Nevertheless, what is clear is that local Seyferts may not, in fact, be appropriate models for high-$z$ quasars, and estimating $M_{\rm BH}$ for large numbers of AGNs that span the full range of observed properties is more complicated than previously assumed.

\subsection{Mapping the BLR}

Further advancements in the quality of reverberation datasets \citep{bentz09,grier12,fausnaugh17,derosa18} have finally begun to allow the initial goal of reverberation mapping -- mapping out the geometry and kinematics of BLR gas -- to  be achieved for a handful of objects \citep{pancoast14,grier17,williams18}.  

While the details vary from object to object, there are similarities as well: the BLR gas is found to be arranged in a thick disk-like configuration at a moderate inclination to our line of sight ($\sim 10-40^{\circ}$).  The bulk motions are generally found to be rotation with varying contributions from inflow.  Furthermore, in these cases $M_{\rm BH}$ may be directly constrained without use of a scaling factor, and in general the values agree well with the constraints based on $\langle f \rangle$ for the same objects.   

Most of the recent analyses have so far focused on H$\beta$, with the exception of the AGN STORM program, which was modeled on the original International AGN Watch effort.  Coordinated space- and ground-based spectroscopic and photometric monitoring, with a backbone of UV spectroscopy provided by \hst, allowed for the most detailed view yet of the BLR and surrounding regions in the AGN NGC\,5548.  Modeling of the reverberation response across the high- and low-ionization emission lines by \citet{williams20} shows that H$\beta$ again arises from a thick disk-like structure.  In this case however, the kinematics show a strong outflowing component, although the orbits may still be bound.  Ly$\alpha$ and \ion{C}{IV}, on the other hand, appear to arise from a shell-like structure, and surprisingly, \ion{C}{IV} appears to have a weaker outflowing contribution than H$\beta$.  

While forward modeling constrains many properties of the BLR gas, the reverberation response may also be treated as an ill-posed inverse problem, thereby making use of all the data including the details that are not adequately fit by models.  \citet{horne21} successfully inverted the AGN STORM data for NGC\,5548, creating the most detailed velocity-delay maps yet recovered, each one being a projected image along axes of isodelay and line-of-sight velocity for the BLR corresponding to a specific emission line.  The velocity-delay maps generally agree with the constraints derived through forward modeling by \citet{williams20}, although a disk-like rather than shell-like geometry is preferred for \ion{C}{IV} and Ly$\alpha$.   There is also evidence in the reverberation response for the presence of an azimuthal structure orbiting on the far side of the \ion{C}{IV} and Ly$\alpha$ BLR during the monitoring period, as has been seen in several double-peaked AGNs \citep{gezari07b,lewis10,schimoia15,schimoia17}.  

An unexpected finding during the AGN STORM campaign was a period where all the emission lines, high-ionization absorption lines and longer wavelength continuum variations decoupled from the UV continuum variability \citep[the so-called ``BLR holiday'';][]{ goad16,goad19} for approximately $60 - 70$ days -- a breakdown of one of the fundamental assumptions of reverberation mapping.   It may be that a change in the ionizing SED incident on the BLR led to the anomalous line behavior \citep{mathur17,goad19},  likely caused by a disk wind \citep{dehghanian20}.  This complication led to unexpected difficulties in the forward modeling and the velocity-delay map inversion analyses, however it also led to intense scrutiny of all the available information within this rich data set.  Thus, the AGN STORM study of NGC\,5548 was also the first reverberation experiment to explore {\it absorption} line reverberation \citep{kriss19,dehghanian19}, which provides a separate probe of the ionizing continuum and the physical conditions of the surrounding photoionized gas.  

\section{Dust reverberation and beyond}  \label{sec:dust}

Near-IR emission in AGNs at about $2-10 \mu$m is thought to be produced by thermal radiation from hot dust \citep[e.g.,][and references therein]{landt11,landt19}.  Since dust only survives at temperatures below the sublimation temperature (in the range 1300 -- 2000 K, depending on the type of dust grains), this determines the inner radius of the dusty torus.  The dust will be illuminated by, and absorb radiation from, the inner X-ray and UV/optical emitting regions, and so variability at these shorter wavelengths will drive variability in the re-radiated dust emission.  The blackbody emission from this hot dust should peak close to the $K$ band, meaning that the inner edge of the dusty torus can be measured through reverberation mapping in the near-IR.  The larger size scale of the dusty torus compared to the disk and BLR leads to longer lags and slower and weaker variability, affording lower cadence monitoring but requiring longer campaigns.  

Early efforts showed the dusty torus to be $\sim$light-year in Fairall 9 \citep{clavel89}, with dust reverberation of 10 AGNs establishing that the dust lags follow a radius-luminosity relation following approximately $R \propto L^{1/2}$, as expected for dust sublimation \citep{okny_horne01}, although the typical radius of the inner wall of the torus is a factor of 2-3 smaller than was initially expected for a population of dust grains with properties similar to the ISM (e.g., \citealt{kishimoto07}).  Long baseline interferometry in the near-infrared has confirmed these small sizes (see \citealt{burtscher16} for a summary), which are now understood to be the result of a population of large graphite grains making up the inner torus wall, as they are the grains that are most resistant to sublimation \citep{garciagonzalez17,honig17}.  

\begin{figure}
\centering
\includegraphics[width=0.7\textwidth]{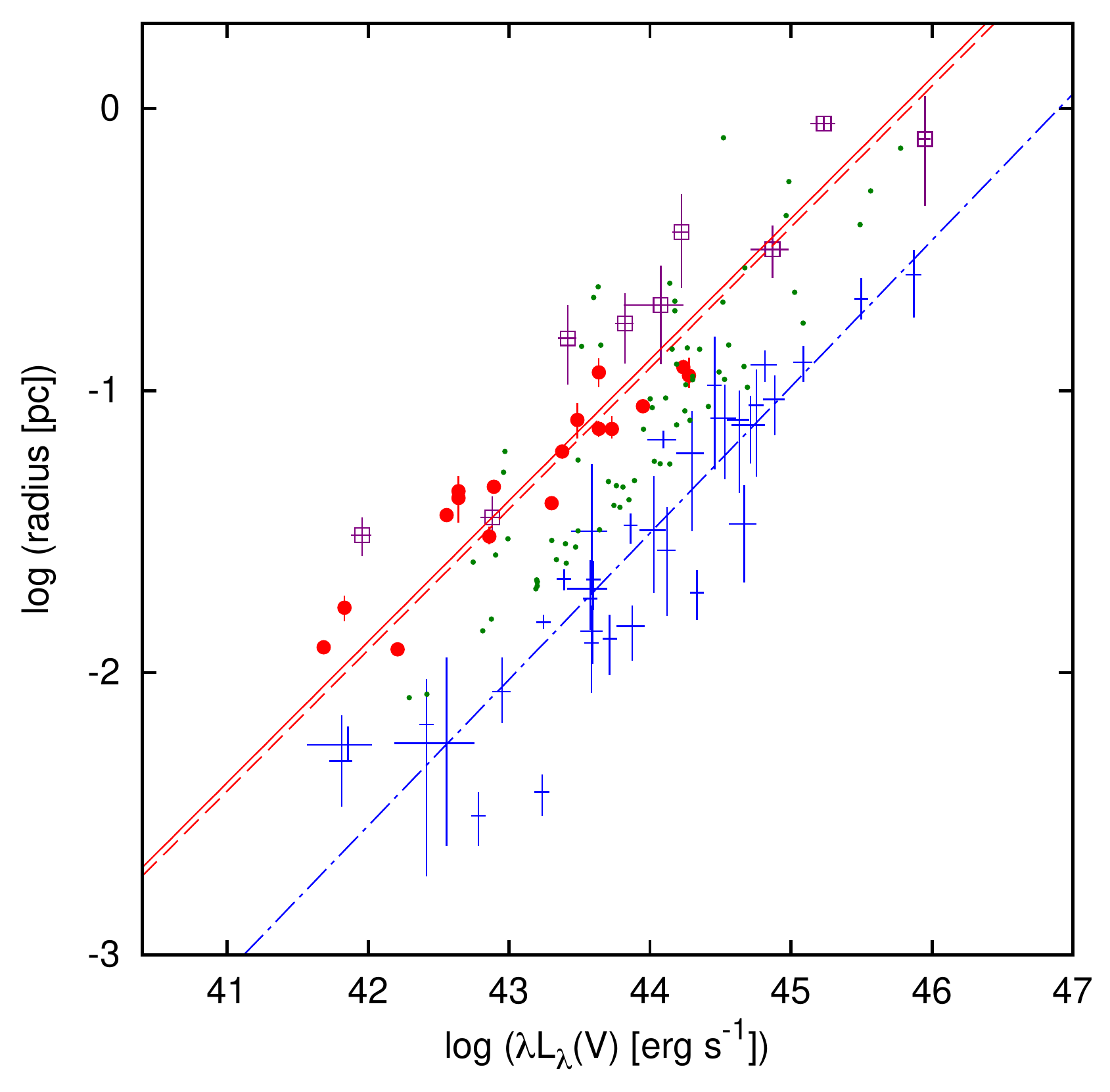}
\caption{A comparison of dust and BLR radii from \citet{koshida14} Red circles indicate $K$-band dust reverberation radii from \citet{koshida14}, purple open squares indicate $K$-band interferometric radii \citep{kishimoto11,weigelt12}, green dots at dust radii from SED fitting \citep{mor12} and blue error bars indicate BLR reverberation radii from \citet{bentz09}. Dust radii are approximately a factor of 4 larger than the BLR radii.}
\label{fig:dustradii}
\end{figure}

In the last 20 years a number of dedicated dust reverberation campaigns have substantially increased the sample of lags \citep[e.g.,][]{suganuma06, koshida14, minezaki19}, as have surveys with repeated imaging that search for transient phenomena \citep{lyu19,yang20}, firmly establishing the dust $R-L$ relation over 4 orders of magnitude in luminosity.  On average the dust lags are $\sim$4 times longer than H$\beta$ lags (e.g., \citealt{koshida14,netzer15}; see Fig.~\ref{fig:dustradii} for a comparison of dust and BLR radii), demonstrating that the bulk of the BLR is enclosed within the dusty torus, as expected from AGN unification arguments. 

The observable effects of central AGNs are expected to extend far beyond the dust torus. More than 20 years of spectrophotometric monitoring of NGC\,5548 enabled the first and only measurement of {\it narrow} emission line reverberation thus far \citep{peterson13}, constraining the [\ion{O}{III}] $\lambda \lambda$\,4959,\,5007 emitting region to $1-3$\,pc in radius, with a light crossing time of approximately a decade.  The timescales required to probe similar and larger structures are generally prohibitive when compared to the lengths of human lifetimes let alone careers.  But just as light echoes from long-dead supernovae may still be found and studied within the Milky Way and its closest companions \citep[e.g.,][]{rest08a,rest08b}, there are indications that distant echoes from AGNs may still be found and studied today, as has been suggested for Hanny's Voorwerp \citep{lintott09}. 

\section{Summary}

The development of reverberation mapping has allowed AGN studies to go beyond the limit of spatial resolution, bringing into view the central regions around supermassive black holes.  The first use of reverberation mapping involved the broad optical and UV emission lines, and BLR reverberation has had a significant impact on measuring black hole masses.  Recent campaigns have started to go to the next stage of mapping out the kinematics and geometry of the BLR.  Further out, reverberation in the near-IR has set the size scale for the inner edge of the dusty torus. The application of reverberation mapping techniques to UV/optical continuum bands, especially with intensive monitoring campaigns led by {\it Swift}, has begun to probe properties of the UV/optical accretion disk but has many open questions relating to size of accretion disks, continuum emission from the BLR and the relationship between the X-ray and UV/optical emitting regions. The development of X-ray reverberation mapping gives us a view to the inner few gravitational radii, close to the black hole, putting constraints on the size scale of the X-ray emitting region, closest to the ISCO.  Putting all these techniques together covers the inner few hundred thousand gravitational radii from light-seconds size scales in the X-rays to light-years in near-IR.  

Looking to the future, the combination of results from large surveys along with focused, intensive multi-wavelength campaigns on individual objects will continue to shape our view of the inner workings on AGNs. Currently ongoing is a large, coordinated X-ray through near-IR monitoring campaign on Mrk~817 built around \hst\ UV spectroscopy, which poses the first opportunity to simultaneously apply all these techniques to the same object. The 10-year Legacy Survey of Space and Time to be performed by the upcoming Vera C. Rubin Observatory\footnote{https://www.lsst.org/} should greatly expand the number AGNs with multiband continuum light curves on a $\sim3$-day cadence, with some fields visited more frequently. Just beginning is also the 5-year SDSS-V\footnote{https://www.sdss5.org/} project, of which a key part is the Black Hole Mapper program, where multi-object spectroscopy of quasars will be used for reverberation mapping. As the field of reverberation mapping enters a new phase, marked by further improved data quality and the application of capable new analysis tools, we are certain to find that new surprises await discovery.

\section{Acknowledgements}
EMC gratefully acknowledges support from the NSF through grant AST-1909199. MCB gratefully acknowledges support from the NSF through grant AST-2009230.  EK acknowledges support from NASA through grant 80NSSC17K0515.
 
\section{Author Contributions}
EMC led Sections 1, 3 and 6. EK led Section 2. MCB led Section 4. EMC and MB co-led Section 5. All authors reviewed and edited the article.

\bibliography{agn}

\end{document}